
\font\bigbf=cmbx10 scaled 1400
\hfuzz 25pt
\hsize=6.1truein
\hoffset=0.2truein

\def\go{\mathrel{\raise.3ex\hbox{$>$}\mkern-14mu
             \lower0.6ex\hbox{$\sim$}}}
\def\lo{\mathrel{\raise.3ex\hbox{$<$}\mkern-14mu
             \lower0.6ex\hbox{$\sim$}}}
\def\ci {{\cal I}}
\def\cc {{\cal C}}

\def\cF {{\cal F}}
\def\Ib5{{\hbox{\rlap{\hbox{\raise.27ex\hbox{-}}}I}}^{(5)}}
\def\bfe {{\bf e}}
\def\bone {{\bar 1}}
\def\btwo {{\bar 2}}
\def\bthr {{\bar 3}}
\def\bi {{\bar i}}
\def\bj {{\bar j}}

\centerline{\bigbf HYDRODYNAMICS OF COALESCING BINARY}
\medskip
\centerline{\bigbf NEUTRON STARS: ELLIPSOIDAL TREATMENT}
\bigskip

\centerline{
DONG LAI$\,$\footnote{$^1$}{Address after September 1, 1994:
Theoretical Astrophysics, California Institute of Technology,
Pasadena, CA 91125}
and $\,$STUART L.~SHAPIRO$\,$\footnote{$^2$}{Departments of Astronomy
and Physics, Cornell University.}}

\medskip
\centerline{\it Center for Radiophysics and Space Research,
Space Sciences Building}
\centerline{\it Cornell University, Ithaca, NY 14853}

\medskip
\centerline{E-mail: dong@astrosun.tn.cornell.edu ~~~
		shapiro@astrosun.tn.cornell.edu}

\bigskip
\bigskip
\centerline{\bf ABSTRACT}
\medskip

We employ an approximate treatment of dissipative hydrodynamics
in three dimensions to study the coalescence of binary neutron
stars driven by the emission of gravitational waves. The stars
are modeled as compressible ellipsoids obeying a polytropic
equation of state; all internal fluid velocities are assumed to
be linear functions of the coordinates. The hydrodynamic equations
then reduce to a set of coupled ordinary differential equations
for the evolution of the principal axes of the ellipsoids, the
internal velocity parameters and the binary orbital parameters.
Gravitational radiation reaction and viscous dissipation are both
incorporated.  We set up exact initial binary equilibrium
configurations and follow the transition from the quasi-static,
secular decay of the orbit at large separation to the rapid
dynamical evolution of the configurations just prior to contact.
A hydrodynamical instability resulting from tidal interactions
significantly accelerates the coalescence at small separation,
leading to appreciable radial infall velocity and tidal lag angles
near contact. This behavior is reflected in the gravitational
waveforms and may be observable by gravitational wave detectors
under construction.

\bigskip
{\it Subject headings:} hydrodynamics --- instabilities --- stars: neutron
--- stars: rotation --- stars: binaries: close
--- radiation mechanisms: gravitational

\bigskip
\bigskip
\centerline{\it Submitted to ApJ, July 1994}

\vfil\eject
\bigskip
\centerline{\bf 1. INTRODUCTION}
\medskip

Coalescing neutron star binaries are the primary targets for
the detection of gravitational waves by the
planed LIGO/VIRGO laser-interferometer system
(Abramovici et al.~1992). The event rate of binary coalescence
has been estimated to be $\sim 10^{2}$ yr$^{-1}$ Gpc$^{-3}$
(Narayan, Piran \& Shemi 1991, Phinney 1991). Extracting
gravity wave signals from noise requires
accurate theoretical waveforms in the frequency range $10-1000$ Hz,
corresponding to the last few minutes of the binaries' life
and orbital separations less than about $700$ km (Cutler et al.~1993).

To leading order, the binary inspiral and the resulting waveform are described
by Newtonian dynamics of two point masses, together with
the lowest-order dissipative effect corresponding to the emission of
gravitational radiation via the quadrupole formula.
One important correction is the effect of post-Newtonian terms,
which can produce large cumulative orbital phase error
(e.g., Cutler et al.~1993), accelerate binary coalescence at small separation
(Lincohn \& Will 1990; Kidder, Will \& Wiseman 1992, 1993),
and cause precession of the orbital plane (Apostolatos et al.~1994).
The other corrections come from hydrodynamical effects due to
the finite size of neutron stars.
The analysis of Bildsten \& Cutler (1992) showed that
the binary cannot be synchronized by viscous torque during the orbital decay
(see also Kochanek 1992). Thus Newtonian spin-orbit coupling has negligible
effect on the orbital phase at large separation, unless the neutron stars have
intrinsic spins near the break-up limit (see also Lai, Rasio \& Shapiro 1994a,
hereafter LRS3). Another aspect of hydrodynamical binary interactions concerns
resonant tidal excitations of g-modes in the stars, which
occur at large orbital separation (orbital frequency $\lo 100$ Hz).
It was shown (Reisenegger \& Goldreich 1994; Lai 1994), however,
that the ``resonant tides'' and their effects on the orbital decay rate are
rather small. Therefore, it has now become clear that
hydrodynamical interactions are important only during the final
stage of neutron star binary coalescence, when the orbital separation
is within a few stellar radii. The importance of tidal effects
havs been demonstrated in our previous studies (LRS3;
Lai, Rasio \& Shapiro 1993b, hereafter LRS2), where it was shown that
close binary systems containing sufficiently incompressible
fluid, such as binary neutron stars, are
dynamically unstable as a result of strong Newtonian
tidal interactions. The basic consequence of the instability is the
acceleration of the coalescence, leading to a significant radial
infall velocity at contact, comparable to the free-fall velocity.

At large orbital separation, before the stability limit is reached,
the neutron star binary evolves quasi-statically along an equilibrium
sequence. This phase of the evolution has been studied in detail in LRS3,
where we have considered the effects of varying neutron star masses, radii,
spins and the stiffness of the equation of state. We have also shown that
stable mass transfer is nearly impossible.
At small orbital separation, the quasi-equilibrium
treatment of LRS3 can provide qualitative features of the binary evolution.
However, to obtain quantitatively accurate results
when the orbital evolution takes place on a timescale comparable to the
internal hydrodynamic timescale, especially when a dynamical
instability is encountered, we need to use fully dynamical equations
to describe the system. A new approximate energy variational formalism
based on compressible ellipsoidal figures has been developed recently
(Lai, Rasio \& Shapiro 1994c, hereafter LRS5) to handle
binary hydrodynamics. The essence of our treatment is to
replace the infinite number of degrees of freedom in a fluid binary
by a small number of variables specifying the essential geometric
and kinematic properties of the system, such as ellipsoid axes,
angular velocity and vorticity. The hydrodynamics is then described
approximately by a set of ordinary differential equations (ODEs)
for the time evolution of these variables,
in place of the usual hydrodynamical PDEs.
In this paper we apply this dynamical theory to neutron star binary
evolution just prior to the final merging.

Our dynamical ellipsoid binary model represents the compressible
generalization of the Riemann-Lebovitz equations
for an isolated, incompressible ellipsoid
(Lebovitz 1966; see also Chandrasekhar 1969, hereafter Ch69).
Our approximation scheme
provides an exact description of Newtonian binary stars when (a) the
fluid is incompressible (polytropic index $n=0$), and (b) the tidal
effects beyond quadrupole order can be ignored. Hence for
neutron stars governed by a moderately stiff equation of state
($n\lo 0.5$) and separated by a few stellar radii,
our model is an excellent approximation
to the true solution. Although our dynamical model is
essentially equivalent to the affine model developed by
Carter \& Luminet (1983, 1986) in the context of tidal interactions with
a massive black hole, our formulation of the problem is quite different
and makes explicit use of global quantities
such as the total angular momentum and fluid circulation, which are conserved
in the absence of dissipation. As a result, appreciable simplification
in the description of many dynamical processes can be achieved.
Moreover, we incorporate both viscosity and gravitational radiation reaction
as possible dissipation mechanisms.

Complete understanding the final coalescence and merger of the neutron
stars requires full 3D hydrodynamical simulations.
So far, all simulations start
from a binary configuration near contact, by which point
hydrodynamical effects are already important
(Oohara \& Nakamura 1990; Nakamura \& Oohara 1991;
Shibata, Nakamura \& Oohara 1992, 1993;
Rasio \& Shapiro 1992,~1994; Davies et al.~1994).
Modeling the distant, pre-contact phase of binary coalescence by
3D numerical codes
would require prohibitively large amounts of computer resources.
Our formulation of binary dynamics in terms of ODEs allows us to
follow the evolution of the system over a large number
of orbits without having to worry about excessive computational
time or about the growth of numerical errors.
Thus our dynamical binary model is particularly useful in
studying the pre-contact transition phase in which the binary evolves from
the quasi-static secular regime to the fully dynamical regime.
In addition, our model can provide reliable initial data for 3D simulations of
the binary merger and can serve as a check of more sophisticated numerical
routines.

The main purpose of this paper is to present the complete dynamical equations
for two stars modeled as ellipsoids in binary orbit (\S 2),
including dissipative forces of viscosity and gravitational radiation reaction
(\S 3). The equations are used to demonstrate the dynamical instability
in the binary (\S 4) and to study a few selected scenarios for neutron star
binary coalescence driven by gravitational radiation (\S 5).
No attempt is made at a complete survey of parameter space. Instead,
we show how straightforward it is to vary the individual masses,
equation of state and spins of the interacting stars so that our method can be
easily employed by future investigators to examine other cases of interest.

We adopt geometrized units and set $G=c=1$ throughout the paper.

\bigskip
\centerline{\bf 2. DYNAMICAL EQUATIONS OF DARWIN-RIEMANN BINARIES}
\nobreak
\medskip

The dynamical equations for compressible Roche-Riemann binaries, consisting of
a finite-size star and a point mass, have been
derived in LRS5. The readers are referred to that paper for the basic
definitions of variables, notation and further details.
The generalization to compressible {\it Darwin-Riemann binaries},
consisting of two finite-size stars, is straightforward and is sketched below.

Consider a binary system containing two stars of mass $M$
and $M'$, each obeying a polytropic equation of state.
Throughout this paper unprimed quantities refer to the star of mass
$M$ and primed quantities refer to the star of mass $M'$.
The density and pressure are related by the relation
$$P=K\rho^{1+1/n},~~~~P'=K'\rho'^{1+1/n'}.\eqno(2.1)$$
Note that for given $n$ and $n'$, the values of $K$ and $K'$ are determined
from the equilibrium radii $R_o$ and $R_o'$ of the
nonrotating stars with the same masses.
Thus any realistic equation of state can be approximately fit
by a polytropic law (cf.~LRS3, \S 4.1).

In the Darwin-Riemann binary model, both stars have the structure
of compressible {\it Riemann-S ellipsoids} (Ch69,
Lai, Rasio \& Shapiro 1993a, hereafter LRS1). A general
Riemann-S ellipsoid is characterized by the angular velocity
${\bf\Omega}=\Omega{\bf e}_z$ {\it of the ellipsoidal figure} (the pattern
speed) about a principal axis and the internal motion of the
fluid with uniform vorticity ${\bf\zeta}=\zeta{\bf e}_z$ along the same axis
(in the frame corotating with the figure).
We assume that the surfaces of constant density inside the star
form self-similar ellipsoids. The number of degrees of freedom for star $M$
is then reduced from infinity to five:
three principal axes $a_1$, $a_2$, $a_3$,
and two angles $\phi$, $\psi$, defined such that $d\phi/dt=\Omega$ and
$d\psi/dt=\Lambda$, where $\Lambda=-a_1a_2\zeta/(a_1^2+a_2^2)$.
Similarly, for star $M'$, we have
$a_1'$, $a_2'$, $a_3'$, $\phi'$ and $\psi'$.
The spins are assumed to be aligned parallel to the orbital
angular momentum. The orbit is specified by the orbital separation $r$ and
the true anomaly $\theta$ (see Fig.~1).
Thus, a total of $12$ variables completely specify
the dynamics of the binary system.

The Lagrangian of the system can be written as the sum of the intrinsic
stellar components $L_s$ and $L_s'$, and the orbital
component $L_{orb}$ according to
$$L=L_s+L_s'+L_{orb}.\eqno(2.2)$$
To derive the stellar term $L_s$, let ${\bf e}_1$, ${\bf e}_1$ and ${\bf e}_3$
be the basis unit vectors
along the instantaneous direction of the principal axes of the
ellipsoid $M$, with ${\bf e}_3$ perpendicular to the orbital plane
(the ``{\it body frame}''). In the inertial
frame, the velocity field ${\bf u}$ of the fluid
inside $M$ relative to the center of mass can be written as
$${\bf u}=\left[\left({a_1\over a_2}\Lambda-\Omega\right)x_2{\bf e}_1
+\left(-{a_2\over a_1}\Lambda+\Omega\right)x_1{\bf e}_2\right]
+\left[{\dot a_1\over a_1}x_1{\bf e}_1+{\dot a_2\over a_2}x_2{\bf e}_2+
{\dot a_3\over a_3}x_3{\bf e}_3\right].
\eqno(2.3)$$
The kinetic energy of $M$ relative to its center of mass is then given by
$$T={1\over2}I(\Lambda^2+\Omega^2)
-{2\over5}\kappa_nMa_1a_2\Lambda\Omega
+{1\over 10}\kappa_nM(\dot a_1^2+\dot a_2^2+\dot a_3^3),\eqno(2.4)$$
where $I=\kappa_nM(a_1^2+a_2^2)/5$ and $\kappa_n\le 1$
is a constant of order unity
which depends only on $n$ (see Table I in LRS1).

The internal energy of $M$ is given by
$$U=\int\!\! n {P\over \rho} dm = k_1K\rho_c^{1/n}M.\eqno(2.5)$$
where $k_1$ is another constant depending only on $n$ and
$\rho_c\propto 1/(a_1a_2a_3)$ is the central density.
The self-gravitational potential energy is given by
$$W=-{3\over5-n}{M^2\over R}{\ci\over 2R^2},~~~~{\rm with}~~~~
\ci=A_1a_1^2+A_2a_2^2+A_3a_3^2,\eqno(2.6)$$
where $R\equiv (a_1a_2a_3)^{1/3}$ is the mean radius of the ellipsoid,
and dimensionless index symbols $A_i$ are defined as in Ch69 (\S17).
Therefore we have
$$L_s=T-U-W.\eqno(2.7)$$
Similar expressions for $M'$ can also be derived.
The orbital component is clearly
$$L_{orb}={1\over 2}\mu\dot r^2+{1\over 2}\mu r^2\dot\theta^2-W_i,\eqno(2.8)$$
where $\mu=MM'/(M+M')$ is the reduced mass.
The gravitational interaction energy $W_i$ between $M$ and $M'$ is
given to quadrupole order by
$$\eqalign{
W_i =& -{MM'\over r}-{M'\over 2r^3}[I_{11}(3\cos^2\alpha-1)
+I_{22}(3\sin^2\alpha-1)-I_{33}]\cr
&-{M\over 2r^3}[I_{11}'(3\cos^2\alpha'-1)
+I_{22}'(3\sin^2\alpha'-1)-I_{33}'],\cr
}\eqno(2.9)$$
where $\alpha=\theta-\phi$, $\alpha'=\theta-\phi'$, and
$I_{ij}=\kappa_n Ma_i^2\delta_{ij}/5$, and similarly for $I_{ij}'$.

Given the Lagrangian, the dynamical equations can then be
obtained from the Euler-Lagrange equations
$${d\over dt}{\partial L\over\partial \dot q_i}=
{\partial L\over\partial q_i},
\eqno(2.10)$$
where $\{q_i\}=\{a_i,\phi,\psi,a_i',\phi',\psi',r,\theta\}$.
For $q_i=\phi$, we have
$${dJ_s\over dt}={3M'\over 2r^3}\sin 2\alpha (I_{11}-I_{22})={\cal
N},\eqno(2.11)$$
where $J_s$ is the ``spin'' angular momentum of $M$
$$J_s ={\partial L\over \partial \Omega}=
I\Omega -{2\over5}\kappa_nMa_1a_2\Lambda,\eqno(2.12)$$
and $\cal N$ is the tidal torque on the star. Similarly, for $q_i=\phi'$,
we have
$${dJ_s'\over dt}={3M\over 2r^3}\sin 2\alpha' (I_{11}'-I_{22}')={\cal N'}.
\eqno(2.13)$$
For $q_i=\theta$, equation (2.13) gives
$${dJ_{orb}\over dt}=-{\cal N}-{\cal N'},\eqno(2.14)$$
where $J_{orb}=\mu r^2\dot\theta$ is the orbital angular momentum.
Thus the total angular momentum
$$J=J_s+J_s'+J_{orb},\eqno(2.15)$$
is conserved. Finally, for $q_i=\psi$ and $q_i=\psi'$, we obtain
$${d\cc\over dt}=0,~~~~{d\cc'\over dt}=0,\eqno(2.16)$$
where $\cc$ is the fluid circulation in star $M$
$$\cc ={\partial L\over \partial\Lambda}=
I\Lambda-{2\over5}\kappa_nMa_1a_2\Omega,\eqno(2.17)$$
and similarly for $\cc'$. Thus the fluid circulations in $M$ and
$M'$ are individually conserved.

The complete dynamical equations can be written
in a numerically convenient form as follows:
$$\eqalignno{
&\ddot a_1 =a_1(\Omega^2+\Lambda^2)-2a_2\Omega\Lambda
-{2\pi\over q_n}a_1A_1\bar\rho
+\left({5k_1 \over n\kappa_n}{P_c\over\rho_c}\right){1\over a_1}
+{M'a_1\over r^3}(3\cos^2\alpha-1),                          &\cr
&							     &(2.18)\cr
&\ddot a_2 =a_2(\Omega^2+\Lambda^2)-2a_1\Omega\Lambda
-{2\pi\over q_n}a_2A_2\bar\rho
+\left({5k_1 \over n\kappa_n}{P_c\over\rho_c}\right){1\over a_2}
+{M'a_2\over r^3}(3\sin^2\alpha-1),                          &\cr
&							     &(2.19)\cr
&\ddot a_3 =-{2\pi\over q_n}a_3A_3\bar\rho
+\left({5k_1 \over n\kappa_n}{P_c\over\rho_c}\right){1\over a_3}
-{M'a_3\over r^3},                                           &(2.20)\cr
&\dot\Omega =\left({a_2\over a_1}-{a_1\over a_2}\right)^{-1}
\left[2\left({\Omega\over a_2}+{\Lambda\over a_1}\right)\dot a_1
-2\left({\Omega\over a_1}+{\Lambda\over a_2}\right)\dot a_2
-{3M'\over 2r^3}\left({a_1\over a_2}+{a_2\over a_1}\right)\sin 2\alpha
\right],      				&\cr
& 							     &(2.21)\cr
&\dot\Lambda =\left({a_2\over a_1}-{a_1\over a_2}\right)^{-1}
\left[2\left({\Omega\over a_1}+{\Lambda\over a_2}\right)\dot a_1
-2\left({\Omega\over a_2}+{\Lambda\over a_1}\right)\dot a_2
-{3M'\over r^3}\sin 2\alpha\right],     		     &(2.22)\cr
&\ddot r =r{\dot\theta}^2-{M+M'\over r^2}
-{3\kappa_n\over 10}{M+M'\over r^4}\left[a_1^2(3\cos^2\alpha-1)
+a_2^2(3\sin^2\alpha-1)-a_3^2\right] &\cr
&~~~~ -{3\kappa_n'\over 10}{M+M'\over r^4}\left[a_1'^2(3\cos^2\alpha'-1)
+a_2'^2(3\sin^2\alpha'-1)-a_3'^2\right],		    &(2.23)\cr
&\ddot \theta =-{2\dot r\dot\theta\over r}
-{3\kappa_n\over 10}{M+M'\over r^5}(a_1^2-a_2^2)\sin 2\alpha-
{3\kappa_n'\over 10}{M+M'\over r^5}(a_1'^2-a_2'^2)\sin 2\alpha',&(2.24)\cr
}$$
where $\dot\theta=\Omega_{orb}$, $\dot\phi=\Omega$, and
$q_n\equiv\kappa_n(1-n/5)$. The equations for
$\ddot a_i$, $\dot\Omega'$ and $\dot\Lambda'$ can be similarly written down by
switching unprimed quantities with primed quantities in equations
(2.18)-(2.24). Also, the pressure term $(5k_1P_c)/(n\kappa_n\rho_c)$
can be conveniently expressed in terms of $R_o$, $M$ and
other dynamical variables as
$${5k_1 \over n\kappa_n}{P_c\over\rho_c}
={M\over q_nR_o}\left({R_o\over R}\right)^{3/n},~~~~~~(n\ne 0),
\eqno(2.25)$$
(see LRS5, where the equivalent expressions for the limiting
case of $n=0$ can also be found).

\bigskip
\centerline{\bf 3. GRAVITATIONAL RADIATION REACTION AND}
\smallskip
\centerline{\bf VISCOUS DISSIPATION}
\nobreak
\medskip

Dissipation modifies Euler-Lagrange equations according to
$${d\over dt}{\partial L\over\partial \dot q_i}=
{\partial L\over\partial q_i}+{\cal F}_{q_i},\eqno(3.1)$$
where ${\cal F}_{q_i}$ is the ``generalized'' force associated with variable
$q_i$, and it is defined via the non-conservative dissipation
rate ${\cal W}$ (Rayleigh's dissipation function, cf.~Goldstein 1980) by
${\cal W}={\cal F}_{q_i}\dot q_i$.

\bigskip
\centerline{\bf 3.1 Gravitational Radiation Reaction}
\nobreak
\medskip

The main driving force for neutron star binary coalescence is
gravitational radiation reaction. Consider
the {\it orbital coordinates} $\{\bfe_\bi\}$,
centered at the center-of-mass (CM) of the system with
$\bfe_{\bone}$ along the line joining $MM'$, $\bfe_\bthr$ perpendcular to the
orbital plane, and $\bfe_\btwo$ perpendicular to
$\bfe_\bone$ and $\bfe_\bthr$ (see Fig.~1). This coordinate basis rotates with
angular velocity $\Omega_{orb}$ with respect to an inertial coordinate system.
In the weak-field, slow motion regime of general relativity,
the gravitational radiation emission induces a back-reaction potential
$\Phi_{react}$, which can be written as (Misner, Thorne \& Wheeler 1973):
$$\Phi_{react}={1\over 5}\Ib5_{\bi\bj}x_\bi x_\bj.\eqno(3.2)$$
where $\Ib5_{\bi\bj}$ is the fifth derivative of
the reduced quadrupole moment
tensor of the system {\it projected onto the orbital frame\/}.
The contribution from $M$ to ${\cal W}$ is given by
$$\eqalign{
{\cal W}_M=& -\int_M\!{\bf v}\cdot\nabla\Phi_{react} dm\cr
= &-{2\over 5}\left({1\over 5}\kappa_nM\right )\biggl [
\left(\Ib5_{\bone\bone}\cos^2\alpha+\Ib5_{\btwo\btwo}\sin^2\alpha
-\Ib5_{\bone\btwo}\sin 2\alpha\right)a_1\dot a_1 \cr
&+\left(\Ib5_{\bone\bone}\sin^2\alpha+\Ib5_{\btwo\btwo}\cos^2\alpha
+\Ib5_{\bone\btwo}\sin 2\alpha\right)a_2\dot a_2
+\Ib5_{\bthr\bthr}a_3\dot a_3 \cr
&+\left(\Ib5_{\bone\btwo}\cos 2\alpha
+(\Ib5_{\bone\bone}-\Ib5_{\btwo\btwo}){1\over 2}\sin 2\alpha\right)
(a_1^2-a_2^2)\Omega \biggr ]\cr
&-{2\over 5}M\left(\Ib5_{\bone\bone}r_{cm}\dot r_{cm}
+\Ib5_{\bone\btwo}r_{cm}^2\Omega_{orb}\right ),    \cr
}\eqno(3.3)$$
where ${\bf v}={\bf u}+{\bf u}_{orb}$ is the fluid velocity in $M$,
$r_{cm}$ is the distance from the CM of the system to the CM of $M$.
A similar expression can be written down for $M'$. Therefore
$$\eqalign{
{\cal W}= &{\cal W}_M+{\cal W}_{M'}\cr
= &-{2\over 5}\left({1\over 5}\kappa_nM\right )\biggl [
\left(\Ib5_{\bone\bone}\cos^2\alpha+\Ib5_{\btwo\btwo}\sin^2\alpha
-\Ib5_{\bone\btwo}\sin 2\alpha\right)a_1\dot a_1 \cr
&+\left(\Ib5_{\bone\bone}\sin^2\alpha+\Ib5_{\btwo\btwo}\cos^2\alpha
+\Ib5_{\bone\btwo}\sin 2\alpha\right)a_2\dot a_2
+\Ib5_{\bthr\bthr}a_3\dot a_3 \cr
&+\left(\Ib5_{\bone\btwo}\cos 2\alpha
+(\Ib5_{\bone\bone}-\Ib5_{\btwo\btwo}){1\over 2}\sin 2\alpha\right)
(a_1^2-a_2^2)\Omega \biggr ]\cr
&-{2\over 5}\left({1\over 5}\kappa_n'M'\right )\biggl [
\left(\Ib5_{\bone\bone}\cos^2\alpha'+\Ib5_{\btwo\btwo}\sin^2\alpha'
-\Ib5_{\bone\btwo}\sin 2\alpha'\right)a_1'\dot a_1' \cr
&+\left(\Ib5_{\bone\bone}\sin^2\alpha'+\Ib5_{\btwo\btwo}\cos^2\alpha'
+\Ib5_{\bone\btwo}\sin 2\alpha'\right)a_2'\dot a_2'
+\Ib5_{\bthr\bthr}a_3'\dot a_3' \cr
&+\left(\Ib5_{\bone\btwo}\cos 2\alpha'
+(\Ib5_{\bone\bone}-\Ib5_{\btwo\btwo}){1\over 2}\sin 2\alpha'\right)
(a_1'^2-a_2'^2)\Omega' \biggr ]\cr
&-{2\over 5}\mu\left(\Ib5_{\bone\bone}r\dot r
+\Ib5_{\bone\btwo}r^2\Omega_{orb}\right ).    \cr
}\eqno(3.4)$$

The dissipative forces due to gravitational radiation are then given by
${\cal F}_{q_i}={\partial {\cal W}/\partial \dot q_i}$.
Clearly, ${\cal F}_\psi={\cal F}_{\psi'}=0$,
thus gravitational radiation reaction
conserves $\cc$ and $\cc'$.
With the inclusion of gravitational radiation reaction forces, the
dynamical equations (2.18)-(2.24) for binaries are modified
according to:
$$\eqalignno{
&\ddot a_1=\{\cdots\}-{2\over 5}\left[\Ib5_{\bone\bone}\cos^2\alpha
+\Ib5_{\btwo\btwo}\sin^2\alpha-\Ib5_{\bone\btwo}\sin 2\alpha\right]a_1,
&(3.5)\cr
&\ddot a_2=\{\cdots\}-{2\over 5}\left[\Ib5_{\bone\bone}\sin^2\alpha
+\Ib5_{\btwo\btwo}\cos^2\alpha+\Ib5_{\bone\btwo}\sin 2\alpha\right]a_2,
&(3.6)\cr
&\ddot a_3=\{\cdots\}-{2\over 5}\Ib5_{\bthr\bthr}a_3,
&(3.7)\cr
&\dot\Omega=\left({a_2\over a_1}-{a_1\over a_2}\right)^{-1}
\left[\{\cdots\}+{2\over 5}\left(\Ib5_{\bone\btwo}\cos 2\alpha
+{1\over 2}(\Ib5_{\bone\bone}-\Ib5_{\btwo\btwo})\sin 2\alpha\right)
\left({a_1\over a_2}+{a_2\over a_1}\right)\right], &\cr
& &(3.8)\cr
&\dot\Lambda=\left({a_2\over a_1}-{a_1\over a_2}\right)^{-1}
\left[\{\cdots\}+{4\over 5}\left(\Ib5_{\bone\btwo}\cos 2\alpha
+{1\over 2}(\Ib5_{\bone\bone}-\Ib5_{\btwo\btwo})\sin 2\alpha\right)\right],
&(3.9)\cr
&\ddot r=\{\cdots\}-{2\over 5}\Ib5_{\bone\bone}r,
&(3.10)\cr
&\ddot \theta=\{\cdots\}-{2\over 5}\Ib5_{\bone\btwo},
&(3.11)\cr
}$$
where $\{\cdots\}$ denote the nondissipative terms that already
exist in equations (2.18)--(2.24) (This notation will be used throughout
the paper). The expressions for $\ddot a_i'$,
$\dot\Omega'$ and $\dot\Lambda'$ are similar.

When the timescale for both orbital and internal quantities to change
is much longer than the rotation period, e.g.,
$|da_i/dt|<<|\Omega a_i|$, simple expressions for $\Ib5_{ij}$ can be
derived. To order $da_i/dt$, the nontrivial components of $\Ib5_{\bi\bj}$ are
$$\eqalign{
\Ib5_{\bone\bone}=& -\Ib5_{\btwo\btwo}
= 16\Omega^5(I_{11}-I_{22})\sin 2\alpha
+16\Omega'^5(I_{11}'-I_{22}')\sin 2\alpha'\cr
&+40\Omega_{orb}^3[2\Omega_{orb}\mu r\dot r+2\dot\Omega \mu r^2]
+40\Omega^3[\Omega(\dot I_{11}-\dot I_{22})+2\dot\Omega
(I_{11}-I_{22})]\cos 2\alpha\cr
&+40\Omega'^3[\Omega'(\dot I_{11}'-\dot I_{22}')+2\dot\Omega'
(I_{11}'-I_{22}')]\cos 2\alpha',\cr
\Ib5_{\bone\btwo}=& \Ib5_{\btwo\bone}
= 16\Omega_{orb}^5\mu r^2+16\Omega^5(I_{11}-I_{22})\cos 2\alpha
+16\Omega'^5(I_{11}'-I_{22}')\cos 2\alpha'\cr
&-40\Omega^3[\Omega (\dot I_{11}-\dot I_{22})
+2\dot\Omega (I_{11}-I_{22})]\sin 2\alpha \cr
& -40\Omega'^3[\Omega' (\dot I_{11}'-\dot I_{22}')
+2\dot\Omega' (I_{11}'-I_{22}')]\sin 2\alpha'.\cr
}\eqno(3.12)$$
These generalize the expressions derived in
Appendix A of LRS5 for Roche-Riemann binaries.

\bigskip
\centerline{\bf 3.2 Viscous Dissipation}
\nobreak
\medskip

Viscous dissipation forces can be easily incorporated into our
dynamical equations. Since the motion of the center of mass of the star
is not affected by viscous dissipation
(which depends only on the shear stresses inside the star),
the viscous forces for an isolated star as derived in LRS5 (\S 4.1)
can be directly applied to binaries.
The dissipation rate due to shear viscosity is
${\cal W}={\cal W}_M+{\cal W}_{M'}$, with
$$\eqalign{
{\cal W}_M= &-{4\over 3}\bar\nu M\biggl [\left({\dot a_1\over a_1}\right)^2
+\left({\dot a_2\over a_2}\right)^2+\left({\dot a_3\over a_3}\right)^2
-\left({\dot a_1\over a_1}\right)\left({\dot a_2\over a_2}\right)
-\left({\dot a_1\over a_1}\right)\left({\dot a_3\over a_3}\right)\cr
&-\left({\dot a_2\over a_2}\right)\left({\dot a_3\over a_3}\right)
\biggr ]-\bar\nu M\Lambda^2\left({a_1^2-a_2^2\over a_1a_2}\right)^2,
}\eqno(3.13)$$
where $\bar\nu$ is the mass-averaged shear kinematic viscosity
$\bar\nu=\int\!\nu\,dm/M$. The expression for ${\cal W}_{M'}$
can be written down similarly.
Since ${\cal W}$ is quadratic in $\dot q_i$,
the dissipative forces are given by
$\cF_{q_i}=(1/2)\partial {\cal W}/\partial\dot q_i$.
Clearly, in the presence of viscosity, the fluid circulations
$\cc$ and $\cc'$ are not conserved. However,
viscous forces do not affect angular momentum, i.e.,
$\cF_\theta=\cF_\phi=\cF_{\phi'}=0$.
The dynamical equations
(2.18)-(2.24) are modified according to
$$\eqalignno{
&\ddot a_1=\{\cdots\}-{10\over 3\kappa_n}\bar\nu\left(
{2\dot a_1\over a_1}-{\dot a_2\over a_2}-{\dot a_3\over a_3}\right)
{1\over a_1}, &(3.14)\cr
&\ddot a_2=\{\cdots\}-{10\over 3\kappa_n}\bar\nu\left(
{2\dot a_2\over a_2}-{\dot a_1\over a_1}-{\dot a_3\over a_3}\right)
{1\over a_2}, &(3.15)\cr
&\ddot a_3=\{\cdots\}-{10\over 3\kappa_n}\bar\nu\left(
{2\dot a_3\over a_3}-{\dot a_1\over a_1}-{\dot a_2\over a_2}\right)
{1\over a_3}, &(3.16)\cr
&\dot\Omega =\left({a_2\over a_1}-{a_1\over a_2}\right)^{-1}
\left[\{\cdots\}+{10\over\kappa_n}\bar\nu{a_1^2-a_2^2\over a_1^2a_2^2}\Lambda
\right],  &(3.17)\cr
&\dot\Lambda =\left({a_2\over a_1}-{a_1\over a_2}\right)^{-1}
\left[\{\cdots\}+{5\over\kappa_n}\bar\nu{a_1^2-a_2^2\over a_1a_2}
\left({1\over a_1^2}+{1\over a_2^2}\right)\Lambda\right], &(3.18)\cr
}$$
and the expressions for $\ddot a_i'$, $\dot\Omega'$ and $\dot\Lambda'$ are
similar, while the expressions for $\ddot r$ and $\ddot\theta$ are
unchanged.

\bigskip
\centerline{\bf 4. DYNAMICAL INSTABILITY IN DARWIN-RIEMANN BINARIES}
\nobreak
\medskip

The dynamical instability in a binary system
results from tidal interactions between the two stars.
The height $h$ of the tidal bulge raised on star $M$
by its companion $M'$ is of order $h\sim R(M'/M)(R/r)^3$.
This tidal deformation makes the gravitational interaction between
$M$ and $M'$ more attractive, and the tidal potential energy is
$$W_{tide}\sim -{M' Q\over r^3} \sim -\kappa_n {M'^2 R^5\over r^6},
\eqno(4.1)$$
(see eq.~[2.9]), where $Q \sim \kappa_n M R h$ is the quadrupole
moment of $M$.
Thus at sufficiently small binary separation, assuming $\kappa_n$ is not too
small (i.e., the star is not too compressible), the binary interaction
potential energy $\sim (-MM'/r+W_{tide})$ becomes steeper than the
point-mass contribution $-MM'/r$. This is the cause
of the dynamical instability in the binary, as is common to all
interaction potentials that are sufficiently steeper than $1/r$
(cf.~Goldstein 1980, \S 3.6).

To determine the dynamical stability limit, one only needs to
construct a sequence of equilibrium models with constant $\cc$ and $\cc'$.
The onset of dynamical instability corresponds to
the turning point in the energy curve along the sequence.
This procedure is described and applied in LRS1 and LRS4,
where numerical values
of the stability limits for various binary models have been tabulated.

Using our dynamical equations with no radiation reaction or viscosity,
we can now study how the instability
develops in time. In Figure~2 we show an example of the time evolution of
an unstable system with
$n=n'=0.5$, $K=K'$, $p=M/M'=1/2$ and $\Lambda=\Lambda'=0$ (corotation).
The dynamical stability limit is at $\hat r=r/(a_1+a_1')=1.174$
(cf.~LRS4, Table~2).
The dynamical equations are integrated numerically using a standard
fifth-order Runge-Kutta scheme with adaptive stepsize (Press et al 1992).
At $t=0$, an unstable equilibrium solution is constructed for
$\hat r=r/(a_1+a_1')=1.17$. This equilibrium solution is then perturbed by
setting $\dot r=-10^{-4}(M/R_o)^{1/2}$. For comparison, the results
of an integration for a stable binary with $\hat r=1.18$, and
with the same applied perturbation are also shown.
We see clearly that as the dynamical instability
develops, $a_1$ increases while $r$ decreases, and this is accompanied
by the significant development of tidal lag angle in the two stars
($\alpha,~\alpha'>0$) and de-synchronization ($\Lambda,~\Lambda>0$).
Of course, the precise evolution of an unstable binary depends on
how the initial configuration is perturbed.

The development of tidal lag in the absence of
fluid dissipation can be qualitatively understood as follows (cf.~Lai 1994).
For star $M$ in the binary system, the tidal potential $\propto 1/r^3$
due to the companion $M'$ acts like an external perturbing force,
with a driving
frequency $\sim \Delta\Omega=\Omega_{orb}-\Omega_s$, where
$\Omega_{orb}$ and $\Omega_s$ are the orbital and spin angular frequencies.
The star has an intrinsic dynamical frequency of order
$\omega_o\sim (M/R_o^2)^{1/2}$.
Schematically, the equation governing the tidal distortion $\xi$
can be written as
$$\ddot\xi+\omega_o^2\xi\propto {1\over r^3}e^{i\Delta\Omega t}.\eqno(4.2)$$
If $r$ and $\Delta\Omega$ were constant in time, then the stationary tide
would be in phase with the driving force. However, when $r$ and/or
$\Omega_{orb}$ change(s) during binary evolution, we have
(assuming $\omega_o>>\Delta\Omega$)
$$\xi\propto {1\over \omega_o^2 r^3(1-i\alpha_{dyn})}e^{i\Delta\Omega t},
\eqno(4.3)$$
where the lag angle is given by
$$\alpha_{dyn}\sim {\Delta\Omega\over\omega_o^2 t_d},\eqno(4.4)$$
and $t_d\sim |r/\dot r|\sim  |\Omega_{orb}/\dot\Omega_{orb}|$ is the
orbital decay timescale. Thus a {\it dynamical tidal lag\/}
arises naturally even without fluid dissipation, and it is due to
the  finite time necessary for the star to
adjust its structure to the rapidly changing tidal potential.
Since the orbit decays rapidly as a result of dynamical instability,
the binary de-synchronizes, and thus $\alpha_{dyn}$ becomes
large at small $r$.

\bigskip
\centerline{\bf 5. NEUTRON STAR BINARY COALESCENCE}
\nobreak
\medskip

The main parameters that enter the
evolution equations of \S 2 and 3 are the mass ratio $p=M/M'$, and the
ratios $R_o/M$ and $R_o'/M'$. The ratio $R_o/M$ is determined from the nuclear
equation of state (EOS). For the canonical neutron star mass
$M=1.4 M_{\odot}$, all EOS's tabulated in Arnett \& Bowers (1977)
give $R_o/M$ in the range of 4--8. Recent microscopic nuclear
calculations indicate that the neutron star
radius $R_o$ is $\simeq 10$ km, almost independent of the mass for
$M$ in the range of $0.8M_{\odot}$ to $1.5M_{\odot}$
(Wiringa, Fiks \& Fabrocini 1988). 
Thus we will use $R_o/M\simeq 5$ as a representative value
and assume $R_o=R_o'$.

A polytrope is only an approximate parametrization for a real EOS.
We can determine the approximate polytropic index $n$ that mimics
the structure of a real neutron star by using the tabulated values
of the moment of inertia of neutron stars (see LRS3, \S 4.1). Typically
we find $n\simeq 0.5$ for $M\simeq 1.4 M_{\odot}$ for the EOS
of Wiringa, Fiks \& Fabrocini (1988).

Other parameters needed for the calculations are the spins
of the two neutron stars $\Omega_s$ and $\Omega_s'$ when the
binary separation is large. For a uniformly rotating neutron star,
the spin rate is limited by mass-shedding,
$\bar\Omega_s\equiv\Omega_s/(M/R_o^3)^{1/2}\lo 0.6$ (Friedman, Ipser
\& Parker 1986, Cook, Shapiro \& Teukolsky 1992).
At large orbital separation, we have $a_1\rightarrow a_2$ and
$\Omega\rightarrow \Omega_{orb}\rightarrow (M+M')^{1/2}/r^{3/2}<<\Omega_s$,
thus $J_s\rightarrow -I\Lambda$ and $\cc\rightarrow I\Lambda$
(see eqs.~[2.12],~[2.17]).
The fluid circulation inside the star, which is conserved during the binary
evolution in the absence of viscosity,
is therefore given by $\cc=-I\Omega_s$. Here we identify
$\Omega_s=-\Lambda(r=\infty)$ as the spin angular velocity at large $r$
(for an axisymmetric spheroid, uniform spin and vorticity are
indistinguishable; cf.~eq.~[2.3]). Note that when $\Omega_s$ is positive
(the spin is in the same direction as the orbital angular momentum), $\cc$
is negative. For simplicity we assume $\Omega_s'=0=\cc'$.

In Figure 3 we show two examples of the pre-contact evolution of
binary neutron stars. Both calculations start at $t=0$ with
$r=5R_o$, and terminate when the surfaces of the stars contact.
The coalescence is driven by gravitational radiation reaction, and
we ignore viscosity for now.
In the first example, the stars have
equal masses, with $R_o/M=R_o'/M'=5$, $n=n'=0.5$, and both have zero spin
$\Omega_s=\Omega_s'=0$. The initial configuration is obtained by solving
the Darwin-Riemann equilibrium equations (see LRS4) with $r/(a_1+a_1')=2.461$.
In the second example, we have $n=n'=0.5$, but now $M/M'=1.2$ so that
$R_o/M=5$ and $R_o'/M'=6$. Also we set
$\Omega_s/(M/R_o^2)^{1/2}=0.4$ (near the maximum
possible value) and $\Omega_s'=0$, corresponding to
$\cc/(M^3R_o)^{1/2}=-0.1204$ and $\cc'=0$. The initial state has
$r/(a_1+a_1')=2.385$ (and $r/R_o=5$), and the nondimensional
vorticity parameters are
$f_R\equiv -(a_1^2+a_2^2)\Lambda/(a_1a_2\Omega)=3.818$, $f_R'=-2$.
Also shown in Figure 3 are the energy curves of the
constant$-\cc$ equilibrium sequences.
Initially, the binary closely follows the equilibrium sequence.
As the dynamical instability develops, the radial velocity increases
considerably, and thereafter the two stars merge hydrodynamically in just
a few orbits. The terminal radial velocity at contact typically reaches
$10\%$ of the free-fall velocity.
This qualitative behavior has already been observed in the simplified
calculations we presented in LRS2 and LRS3, where the stars were
constrained to retain their equilibrium shapes throughout the evolution.

 From Fig.~3, we also see that the tidal lag angle $\alpha$ or $\alpha'$
increases with decreasing $r$, attaining a large value of about $10^\circ$
at binary contact. Let us consider the lag angle $\alpha$ in star $M$.
In the absence of fluid viscosity, there are two
contributions to this lag angle, $\alpha=\alpha_{dyn}+\alpha_{gr}$:
one is the dynamical lag $\alpha_{dyn}$ discussed in \S 4, which
arises from the rapid orbital decay (especially when the dynamical
instability develops); the other is
the {\it gravitational radiation dissipation lag\/} $\alpha_{gr}$
analogous to the usual viscous lag (see later).
During the secular evolution phase (before the dynamical instability
develops), the orbital decay timescale is
$$t_d\simeq \left|{r\over \dot r}\right|\simeq {5r^4\over 64MM'M_t},
\eqno(5.1)$$
where $M_t\equiv M+M'$.
 From equation (4.4), the dynamical lag is given by
$$\alpha_{dyn}\sim {R_o^3\Delta\Omega\over M}
\left({64\,MM'M_t\over 5\,r^4}\right)
\simeq {64\over 5}{R_o^3 M'M_t^{3/2}\over r^{11/2}},\eqno(5.2)$$
where in the second equality we have specialized in the case when
$\Omega_s\simeq 0$ and $\Delta\Omega\simeq \Omega_{orb}\simeq (M_t/r^3)^{1/2}$.
The gravitational radiation dissipation lag arises because
gravitational radiation reaction directly exerts a torque
on the star. From equation (3.4), this torque ${\cal N}_{gr}={\cal F}_\phi
=\partial{\cal W}/\partial\dot\phi$ is given by
$$\eqalign{
{\cal N}_{gr}= &-{2\over 5}\left({1\over 5}\kappa_nM\right)
\left[\Ib5_{\bone\btwo}\cos 2\alpha
+(\Ib5_{\bone\bone}-\Ib5_{\btwo\btwo}){1\over 2}\sin 2\alpha\right]
(a_1^2-a_2^2) \cr
\simeq & -{2\over 5}\left({1\over 5}\kappa_nM\right )
\left(16\Omega_{orb}^5\mu r^2\cos 2\alpha\right)(a_1^2-a_2^2),\cr
}\eqno(5.3)$$
where in the second equality we have used equation (3.12).
Since the fluid circulation $\cc$ in the star is conserved in the absence of
viscosity, and since $|J_s|\simeq |\cc|$ to the leading order (cf.~eqs.~[2.12],
[2.17]), there must be a small misalignment $\alpha_{gr}$ of the tidal bulge
so that ${\cal N}_{gr}$ can be balanced by the tidal torque
${\cal N}$ (cf.~eq.~[2.11]). Requiring $dJ_s/dt={\cal N}+{\cal N}_{gr}=0$,
we obtain
$$\alpha_{gr}\simeq {32\over 15}{ M M_t^{3/2}\over r^{5/2}},\eqno(5.4)$$
which is independent of the radius and spin of the star.
This result was derived previously in LRS5 (\S 9.2) for Roche-Riemann binaries.
Although $\alpha_{gr}$ is larger than $\alpha_{dyn}$ for
large binary separation $r\go 2 R_o(M'/M)^{1/3}$, the radiation dissipation lag
is always small ($\alpha_{gr}\lo 0.01$).
At smaller orbital separation, the dynamical lag dominates.
Moreover, as the dynamical instability develops, the orbital decay time
becomes comparable to the orbital period, $t_d\sim t_{orb}=1/\Omega_{orb}$,
and the thus the lag angle near binary contact is
$$\alpha\simeq \alpha_{dyn}
\sim {\Omega_{orb}\Delta\Omega\over\omega_o^2}
\sim \left({\Omega_{orb}\over\omega_o}\right)^2
\sim \left({M_t\over M}\right)^2\left({R_o\over r}\right)^3.
\eqno(5.5)$$
For $M\sim M'$ and $r\simeq 2.6R_o$ (contact), we have $\alpha\sim 0.2$,
in agreement with the numerical results of Fig.~3.
Equation (5.5) implies that $\alpha$ is smaller for the spinning,
more massive star, also in agreement with Fig.~3.

The gravitational wave forms can also be calculated using the standard
quadrupole formula, giving
$$\eqalign{
h_{+} &=-{2\over D} \left[\mu r^2\Omega_{orb}^2\cos 2\theta
+(I_{11}-I_{22})\Omega^2\cos 2\phi
+(I_{11}'-I_{22}')\Omega'^2\cos 2\phi'\right](1+\cos^2\Theta),\cr
h_{\times} &=-{4\over D} \left[\mu r^2\Omega_{orb}^2\sin 2\theta
+(I_{11}-I_{22})\Omega^2\sin 2\phi
+(I_{11}'-I_{22}')\Omega'^2\sin 2\phi'\right]\cos\Theta,\cr
}\eqno(5.6)$$
where $D$ is the source distance, and $\Theta$ specifies the angle
between the direction of wave propagation and the $z$-axis.
We have neglected higher-order terms which are smaller by a factor
of order $|\dot r/r\Omega_{orb}|$.
Equation (5.6) generalizes the expressions derived in LRS3
to the case when $\phi\neq \theta$ and $\phi'\neq\theta$ (cf.~Fig.~1).
Neglecting the small tidal correction,
the wave frequency is $f\simeq (M+M')^{1/2}r^{-3/2}/\pi
=1123~M_{1.4}^{-1}(R_o/5M)^{-3/2}(r/3R_o)^{-3/2}$ Hz (for $M=M'$).
Figure 4 depicts
the wave amplitude seen by an observer along the rotation axis ($\Theta=0$)
for one of the cases ($\Omega_s=\Omega_s'=0$) shown in Figure 3.
Comparing to the result for two-point masses, we see that
a large phase error ($2-3$ cycles) develops during the last $\sim 10$
wave cycles as the result of the accelerated coalescence
induced by the dynamical instability.
Such signature should be detectable by specially configured
(``dual recycled'') interferometers that operate over adjustable, narrow
bands around $1000$ Hz (e.g., Strain \& Meers 1991).

The dynamical effects of viscous dissipation can also be incorporated in
the calculations. Dimensionally, the maximum possible value of
viscosity is $\bar\nu_{max}\sim (MR_o)^{1/2}$. However,
the viscosity due to electron-electron scattering
(Flowers \& Itoh 1979), which is the dominant source of microscopic
viscosity since neutrons and protons are likely to be
in superfluid states in cold coalescing neutron stars,
is many orders of magnitude smaller than
$\bar\nu_{max}$. Nevertheless, to illustrate the qualitative effects,
we consider an extreme example, adopting $\bar\nu=0.5R_o(M/R_o)^{1/2}$.
In Figure 5, we compare neutron star binary evolution with and without
viscosity. We see that viscous dissipation tends to synchronize
the binary, i.e., to reduce $\Lambda$ as compared to the inviscid cases.
Such synchronization is the result of viscous tidal lag
$\alpha_{vis}\sim \Delta\Omega/(\omega_o^2 t_{vis})$ (compare with
eq.~[4.4]), where $t_{vis}\sim R_o^2/\bar\nu$ is the viscous timescale
(e.g., Zahn 1977; also see eq.~[8.4] in LRS5).
Also viscous dissipation tends to accelerate the
coalescence. This is because orbital angular momentum is
transferred to the stellar spin via viscous torque.
However, as can be seen In Fig.~5(c),
even with such an extreme value of viscosity,
synchronization can never be achieved. In fact the binary always becomes
more and more asynchronized as $r$ decreases, even if it is corotating
at large separation. This conclusion agrees with that
of Bildsten \& Cutler (1992).

Finally, it should be noted that
our treatment so far has ignored post-Newtonian (PN) effects other than
the lowest-order dissipative effect corresponding to the emission of
gravitational radiation.
However, for the typical value of $R_o/M=5$, other PN effects are
important and can alter the orbits considerably. In particular,
even in the case of two point masses, these PN effects can
by themselves make a circular
orbit become unstable when the separation is smaller than some critical value
(``inner-most stable orbit'')
$r_{GR}$. Kidder, Will \& Wiseman (1992)  have recently obtained
$r_{GR} \simeq 6(M+M')+4\mu$. The PN effects
lead to a plunge orbit for $r<r_{GR}$, with significant infall radial
velocity. For two point masses with $M=M'$,
$r_{GR} \simeq 14 M$. Since the Newtonian hydrodynamical stability limit
is at $r_m \simeq 3 R_o$ typically, $r_{GR}$ and $r_m$ are comparable
for $R_o/M=5$. Clearly, without including PN terms, our
final coalescence trajectory can only be approximate. However, it is
clear from our discussion
that the Newtonian hydrodynamic effects discussed in this paper
are at least as important as relativistic corrections
to the orbital motion for the final phase of neutron star binary
coalescence. When the two effects are both properly incorporated,
the final coalescence is likely to be
even faster, and assume a significant ``head-on'' character.

\bigskip
\centerline{\bf 6. CONCLUSIONS}
\nobreak
\medskip

We have presented a simplified hydrodynamical treatment of close binary
systems based on compressible ellipsoids obeying a polytropic equation of
state. We employed this treatment to demonstrate
the development of dynamical instability during the final phase of
neutron star binary coalescence prior to contact. Such instability is
accompanied by a significant radial infall velocity and dynamical
tidal lag angles of order $10^\circ$. We have also shown that
the neutron star becomes more asynchronized as the binary orbit shrinks
(see also Bildsten \& Cutler 1992). Full hydrodynamical simulations
of binary merger based on non-corotating models with
large initial plunging velocity components (Shibata el al.~1992, 1993) show
significantly different results from simulations
based on corotating binary models without large infall velocities
(Nakamura \& Oohara 1991, Rasio \& Shapiro 1992,~1994).
However, these simulations
should be considered only preliminary since the calculations
start near contact, by which time dynamical effects are already
important and the initial conditions in these simulations are very
approximate.

Our dynamical binary models provide a new tool to study the pre-contact
evolution of binary neutron stars. The simplicity of replacing
the full hydrodynamical equations with ODEs allows us
to sample a large number of
physical parameters (e.g., masses, spins, equations of states, etc.)
and incorporate dissipative effects
(gravitational radiation and viscosity) easily.
Our Newtonian dynamical binary equations, when coupled with
the post-Newtonian equations for point-mass binaries (e.g.,
Kidder et al.~1993), should give a reasonable description of
the terminal phase of the binary prior to merger.

The numerical codes implementing the equations presented in this
paper can be obtained from the authors upon request.

\bigskip
\bigskip
\medskip

This work has been supported in part
by NSF Grant AST 91-19475 and NASA Grant NAGW-2364 to Cornell University.

\vfill\eject
\bigskip
\centerline{\bf REFERENCES}
\medskip
\def\bysame{\hbox to 50pt{\leaders\hrule height 2.4pt depth -2pt\hfill .\ }}
\def\hi{\noindent \hangindent=2.5em}

\hi{
Abramovici, A., et al. 1992, Science, 256, 325}

\hi{
Apostolatos, T.A., Cutler, C., Sussman, G.J., \& Thorne, K.S. 1994,
Phy. Rev. D, in press}

\hi{
Arnett, W.D., \& Bowers, R.L. 1977, ApJS, 33, 415}


\hi{
Bildsten, L., \& Cutler, C. 1992, ApJ, 400, 175}

\hi{
Carter, B., \& Luminet, J.P. 1985, MNRAS, 212, 23}

\hi{
Chandrasekhar, S. 1969, Ellipsoidal Figures of Equilibrium
(New Haven: Yale University Press) (Ch69)}

\hi{
Cook, G.B., Shapiro, S.L., \& Teukolsky, S.A. 1992, ApJ, 398, 203}

\hi{
Cutler, C. et al. 1993, Phys. Rev. Lett., 70, 2984}

\hi{
Davies, M.B., Benz, W., Piran, T., \& Thielemann, F.K. 1994, ApJ, in press}

\hi{
Flowers, E., \& Itoh, N. 1979, ApJ, 230, 847}

\hi{
Friedman, J.L., Ipser, J.R., \& Parker, L. 1986, ApJ, 304, 115}

\hi{
Goldstein, H. 1980, Classical Mechanics (Reading: Addison-Wesley)}


\hi{
Kidder, L.E., Will, C.M., \& Wiseman, A.G. 1992,
Class. Quantum Grav., 9, L125}

\hi{
\bysame 1993, Phys. Rev. D, 47, 3281}

\hi{
Kochanek C.S. 1992, ApJ, 398, 234}

\hi{
Lai, D. 1994, MNRAS, in press}

\hi{
Lai, D., Rasio, F.A., \& Shapiro, S.L. 1993a, ApJS, 88, 205 (LRS1)}

\hi{
\bysame 1993b, ApJ, 406, L63 (LRS2)}

\hi{
\bysame 1994a, ApJ, 420, 811 (LRS3)}

\hi{
\bysame 1994b, ApJ, 423, 344 (LRS4)}

\hi{
\bysame 1994c, ApJ, in press (LRS5)}

\hi{
Lebovitz, N.R. 1966, ApJ, 145, 878}

\hi{
Lincoln, W., \& Will, C. 1990, Phys. Rev. D, 42, 1123}

\hi{
Luminet, J.P., \& Carter, B. 1986, ApJS, 61, 219}

\hi{
Misner, C.M., Thorne, K.S., \& Wheeler, J.A. 1973,
Gravitation (New York: Freeman)}

\hi{
Nakamura, T., \& Oohara, K. 1991, Prog. Theor. Phys., 86, 73}

\hi{
Narayan, R., Piran, T., \& Shemi, A. 1991, ApJ, 379, L17}

\hi{
Oohara, K., \& Nakamura, T. 1990, Prog. Theor. Phys., 83, 906}

\hi{
Phinney, E.S. 1991, ApJ, 380, L17}

\hi{
Press, W.~H., Teukolsky, S.~A., Vetterling, W.~T, \& Flannery, B.~P. 1992,
Numerical Recipes: The Art of Scientific Computing, 2nd Ed.
(Cambridge: Cambridge Univ.\ Press)}

\hi{
Rasio, F. A., \& Shapiro, S.L. 1992, ApJ, 401, 226}

\hi{
Rasio, F. A., \& Shapiro, S.L. 1994, Preprint}

\hi{
Reisenegger, A., \& Goldreich, P. 1994, ApJ, 426, 688}

\hi{
Shibata, M., Nakamura, T., \& Oohara, K. 1992, Prog. Theor. Phys., 88, 1079}

\hi{
\bysame 1993, Prog. Theor. Phys., 89, 809}

\hi{
Strain, K.A., \& Meers, B.J. 1991, Phys. Rev. Lett., 66, 1391}

\hi{
Wiringa, R. B., Fiks, V., \& Fabrocini, A. 1988, Phys. Rev. C, 38, 1010}

\hi{
Zahn, J.P. 1977, A\&A, 57, 383}

\vfil\eject
\centerline{\bf Figure Captions}
\bigskip

\noindent
{\bf FIG.~1}.---
The coordinate system used for Darwin-Riemann binaries.

\noindent
{\bf FIG.~2}.---
The development of the dynamical instability in a
binary system with $p=M/M'=1/2$, $n=n'=0.5$, $K=K'$.
The initial configurations are in equilibrium, and corotating.
The left panels show the evolution of a dynamically unstable binary
with $\hat r=r/(a_1+a_1')=1.17$, while the right panels show that of
a stable binary with $\hat r=1.18$. The solid lines correspond
to star $M$, the dashed lines correspond to $M'$.
Each line is labeled by the value of $i$ in $a_i$.

\noindent
{\bf FIG.~3}.---
The evolution of coalescing binaries driven by gravitational radiation.
The system energy $E$, tidal angle $\alpha$, radial velocity $v_r=\dot r$
and time $t$ are shown as a function of binary separation $r$.
Here $n=n'=0.5$, $R_o/M=5$, $R_o=R_o'$, and the calculations start ($t=0$)
at $r/R_o=5$ and terminate when the surfaces of the stars contact.
The left panels show the case
with $M=M'$ and $\Omega_s=\Omega_s'=0$; the right panels
show the case with $M/M'=1.2$,
$\Omega_s/(M/R_o^3)^{1/2}=0.4$ and $\Omega_s'=0$.
The dotted curve in the $E(r)$ diagram is the
equilibrium energy of a constant-$\cc$ binary sequence.
The turning point marks the onset of dynamical instability.
The dashed line in the right panel corresponds to $\alpha'$, while the
solid line corresponds to $\alpha$. 

\noindent
{\bf FIG.~4}.---
The waveform from neutron star binary coalescence shown in Fig.~3
($\Omega_s=\Omega_s'=0$). The dark solid line corresponds to zero
viscosity, the light solid line assumes $\bar\nu=0.5(MR_o)^{1/2}$.
The dotted line is  the result for two point masses.

\noindent
{\bf FIG.~5}.---
Radial velocity $v_r$, number of orbital cycles $N_{orb}$
and vorticity parameter $\Lambda$ (measuring the degree of non-synchronization)
of a coalescing neutron star binary.
Here $M=M'$, $R_o/M=5$ and $n=0.5$. The solid lines
are for $\cc=\cc'=0$ (irrotation) with zero viscosity, the
dashed lines for $\cc=\cc'=0$ at $r/R_o=5$
and with $\bar\nu=0.5(MR_o)^{1/2}$;
the long-dashed lines are for $\Lambda=\Lambda'=0$ (corotation)
at $r/R_o=5$ with no viscosity, and the dotted-dashed lines are for
$\bar\nu=0.5(MR_o)^{1/2}$. The dotted lines in (a)-(b) are the results for two
point masses. 

\end